\documentclass{article}

\usepackage{color}
\usepackage{amsmath,amsthm}
\usepackage{amsfonts,amssymb}
\usepackage{tikz}
\usepackage{tikzfeynman}   

\definecolor{myurlcolor}{rgb}{0.6,0,0}
\definecolor{mycitecolor}{rgb}{0,0,0.8}
\definecolor{myrefcolor}{rgb}{0,0,0.8}
\usepackage[bookmarks=false]{hyperref}
\hypersetup{colorlinks,
linkcolor=myrefcolor,
citecolor=mycitecolor,
urlcolor=myurlcolor}

\newcommand{\bare}{\mathrm{bare}}
\newcommand{\elec}{\mathrm{elec}}
\newcommand{\ren}{\mathrm{ren}}
\newcommand{\ret}{\mathrm{ret}}
\newcommand{\ext}{\mathrm{ext}}
\newcommand{\sing}{\mathrm{sing}}
\newcommand{\reg}{\mathrm{reg}}

\newcommand{\maps}{\colon}

\newcommand{\R}{\mathbb{R}}

\newcommand{\beq}{\begin{equation}}
\newcommand{\eeq}{\end{equation}}

\begin{document}



\begin{center}   
  {\bf Struggles with the Continuum \\}   
  \vspace{0.3cm}
  {\em John\ C.\ Baez \\}
  \vspace{0.3cm}
  {\small
 Department of Mathematics \\
    University of California \\
  Riverside CA, USA 92521 \\ and \\
 Centre for Quantum Technologies  \\
    National University of Singapore \\
    Singapore 117543  \\    } 
  \vspace{0.3cm}   
  {\small email:  baez@math.ucr.edu \\} 
  \vspace{0.3cm}   
  {\small \today}
  \vspace{0.3cm}   
\end{center}   

Is spacetime really a continuum, with points being described---at least locally---by lists of  real numbers?  Or is this description, though immensely successful so far, just an approximation that breaks down at short distances?  Rather than trying to answer this hard question, let us look back at the struggles with the continuum that mathematicians and physicists have had so far.

The worries go back at least to Zeno.  Among other things, he argued that that an arrow can never reach its target:

\begin{quote}
\emph{That which is in locomotion must arrive at the half-way stage before it arrives at the goal}. ---  Aristotle \cite{AristotleZeno1}.
\end{quote}

\noindent
and Achilles can never catch up with a tortoise:

\begin{quote}
\emph{In a race, the quickest runner can never overtake the slowest, since the pursuer must first reach the point whence the pursued started, so that the slower must always hold a lead}. --- Aristotle \cite{AristotleZeno2}.
\end{quote}

These paradoxes can now be dismissed using the theory of convergent sequences:
a sum of infinitely many terms can still converge to a finite answer.  But this theory is far from trivial.  It became fully rigorous long after the rise of Newtonian physics.  At first, the practical tools of calculus seemed to require infinitesimals, which seemed logically suspect.  Thanks to the work of Dedekind, Cauchy, Weierstrass, Cantor and others, a beautiful formalism was developed to handle the concepts of infinity, real numbers, and limits in a precise axiomatic manner. 

However, the logical problems are not gone.  G\"odel's theorems hang like  a dark cloud over the axioms of set theory, assuring us that any consistent theory as strong as Peano arithmetic, or stronger, will leave some questions unsettled.  For example: how many real numbers are there?   The continuum hypothesis proposes a conservative answer, but since this is independent of the usual axioms of set theory, the question remains open: there could be \emph{vastly} more real numbers than most people think.  Worse, the superficially plausible axiom of choice---which amounts to saying that the product of  any collection of nonempty sets is nonempty---has scary consequences, like the existence of nonmeasurable subsets of the real line.  This in turn leads to results like that of
Banach and Tarski: one can partition a ball of unit radius into six disjoint subsets, and by rigid motions reassemble  these subsets into {\it two} disjoint balls of unit radius.  
(One can even do the job with five, but no fewer \cite{Wagon}.)

Most mathematicians and physicists are inured to these logical problems.  Few of us bother to learn about attempts to tackle them head-on, such as:
\begin{itemize}
\item nonstandard analysis and synthetic differential geometry, which let us work consistently with infinitesimals \cite{Kock1,Kock2,LoebWolff,Robinson},
\item constructivism, in which one must `construct' a mathematical object to prove that it exists \cite{Bishop},
\item  finitism (which avoids completed infinities altogether) \cite{Ye},
\item ultrafinitism, which even denies the existence of very large numbers \cite{Boucher}.
\end{itemize}
This sort of foundational work proceeds slowly, and is deeply unfashionable.  One reason is that 
it rarely seems to impact `real life'.   For example, it seems that no question about the \emph{experimental consequences} of physical theories has an answer that depends on whether 
or not we assume the continuum hypothesis or the axiom of choice.  

But even if we take a hard-headed practical attitude and leave logic to the logicians, our struggles with the continuum are far from over.  In fact, the infinitely divisible nature of the real line---the 
existence of arbitrarily small real numbers---is a serious challenge to physics. 

One of the main goals of physics is to find theories that systematically generate
predictions---for example, predictions of the future state of a system given 
knowledge of its present state.   However, even setting aside the question of whether
these predictions are correct, there is a problem.  For many of the most widely used physical theories,  \emph{we have been unable to rigorously prove that they give
predictions in all circumstances}.  Integrals may diverge, differential equations may
fail to have solutions, and so on.   Underlying all these difficulties is a struggle with the continuum nature of spacetime itself.   

One might hope that a radical approach to the foundations of mathematics---such as those listed above---would allow us to sidestep these problems.  However, I know of no significant progress along these lines.  Some of the ideas of constructivism have been embraced by topos theory, which also provides a foundation for calculus with infinitesimals \cite{Kock1,Kock2}.  Topos theory and especially `$\infty$-topos theory' are becoming important in mathematical physics \cite{Schreiber}.  They shed new light on gauge theory, string theory and other topics.  But as far as I know, they have not yet been used to solve, or get around, the problems we discuss here.

Many physicists believe that a successful theory of quantum gravity will dramatically change our concept of spacetime and shed new light on the problems that plague our existing theories.   I, too, am inclined to believe this.   However, at present, no theory of quantum gravity has made a single experimentally verified prediction.  Moreover, these theories are all beset with their own internal problems.  Thus, in this survey, we limit ourselves to theories that apparently \emph{have} been successful in making predictions, and examine the `cracks' in these theories: the problems that arise from assuming spacetime is a continuum.

Let us look at some examples.  

\section{Newtonian gravity}
\label{newtonian}

In its simplest form, Newtonian gravity describes ideal point particles attracting each other with a force inversely proportional to the square of their distance.  It is one of the early triumphs of modern physics.  But what happens when these particles collide?  Apparently the force between them becomes infinite.  What does Newtonian gravity predict then?

Of course real planets are not points: when two planets come too close together, this idealization breaks down.  Yet if we wish to study Newtonian gravity \emph{as a mathematical theory}, we should consider this case.  Part of working with a continuum is successfully dealing with such issues.

In fact, there is a well-defined `best way' to continue the motion of two point masses through a collision.  Their velocity becomes infinite at the moment of collision but is finite before and after.  The total energy, momentum and angular momentum are unchanged by this event.   So, a 2-body collision is not a serious problem.  But what about a simultaneous collision of 3 or more bodies?  This seems more difficult.

Worse than that, Xia proved in 1992 that with 5 or more particles, there are
solutions where particles shoot off to infinity in a finite amount of time \cite{SaariXia,Xia}.   This sounds crazy at first, but it works like this: a pair of heavy particles orbit each other, another pair of heavy particles orbit each other, and these pairs toss a lighter particle back and forth.  Each time they do this, the two pairs move further apart from each other, while the two particles within each pair get closer together.   Each time they toss the lighter particle back and forth, the two pairs move away from each other faster.  As the time $t$ approaches a certain value $T_0$, the speed of these pairs approaches infinity, so they shoot off to infinity in opposite directions in a finite amount of time, and the lighter particle bounces back and forth an infinite number of times.

Of course this isn't possible in the real world, but Newtonian physics has no `speed limit', and we're idealizing the particles as points. So, if two or more of them get arbitrarily close to each other, the potential energy they liberate can give some particles enough kinetic energy to zip off to infinity in a finite amount of time! After that time, the solution is undefined.  

You can think of this as a modern reincarnation of Zeno's paradox.  Suppose you take a coin and put it heads up.  Flip it over after 1/2 a second, and then flip it over after 1/4 of a second, and so on.  After one second, which side will be up?  There is no well-defined answer.  That may not bother us, since this is a contrived scenario that seems physically impossible.  It's a bit more bothersome that Newtonian gravity doesn't tell us what happens to our particles when $t = t_0$.

One might argue that collisions and these more exotic `noncollision singularities' occur with probability zero, because they require finely tuned initial conditions.  If so, perhaps we can safely ignore them.  

This is a nice fallback position.  But to a mathematician, this argument demands proof.   A bit more precisely, we would like to prove that the set of initial conditions for which two or more particles come arbitrarily close to each other within a finite time has `measure zero'.  This would mean that `almost all' solutions are well-defined for all times, in a very precise sense.

In 1977, Saari proved that this is true for 4 or fewer particles \cite{Saari}.  However, to the best of my knowledge, the problem remains open for 5 or more particles.  Thanks to previous work by Saari, we know that the set of initial conditions that lead to collisions has measure zero, regardless of the number of particles \cite{Saari_collision_1,Saari_collision_2}.  So, the main remaining problem is to prove that for 5 or more particles, noncollision singularities occur with probability zero.  For 4 particles, nobody knows if such singularities can occur at all.  For 3 or fewer we know they do not.

It is remarkable that even Newtonian gravity, often considered a prime example of determinism in physics, has not been proved to make definite predictions, not even `almost always'.   In 1840, Laplace \cite{Laplace} wrote:

\begin{quote}
We ought to regard the present state of the universe as the effect of its antecedent state and as the cause of the state that is to follow. An intelligence knowing all the forces acting in nature at a given instant, as well as the momentary positions of all things in the universe, would be able to comprehend in one single formula the motions of the largest bodies as well as the lightest atoms in the world, provided that its intellect were sufficiently powerful to subject all data to analysis; to it nothing would be uncertain, the future as well as the past would be present to its eyes. The perfection that the human mind has been able to give to astronomy affords but a feeble outline of such an intelligence.
\end{quote}

\noindent However, this dream has not yet been realized for Newtonian gravity.  

I expect that noncollision singularities \emph{will} be proved to occur with probability zero.  If so, the remaining question would be why it takes so much work to prove this, and thus prove that Newtonian gravity makes definite predictions in almost all cases.  Is this is a weakness in the theory, or just the way things go?   Clearly it has something to do with three idealizations:
\begin{itemize} 
\item point particles whose distance can be arbitrarily small,
\item potential energies that can be arbitrarily large and negative, and 
\item velocities that can be arbitrarily large.
\end{itemize}  
These are connected: as the distance between point particles approaches zero, their potential energy approaches $-\infty$, and conservation of energy dictates that some velocities approach $+\infty$.  

Does the situation improve when we go to more sophisticated theories?  For example, does the `speed limit' imposed by special relativity help the situation?  Or might quantum mechanics help, since it describes particles as `probability clouds', and puts limits on how accurately we can simultaneously know both their position and momentum?

We begin with quantum mechanics, which indeed does help.

\section{Quantum mechanics of charged particles}
\label{quantum}

Few people spend much time thinking about `quantum celestial mechanics'---that is, quantum particles obeying Schr\"odinger's equation, that attract each other gravitationally, obeying an inverse-square force law.   But Newtonian gravity is a lot like the electrostatic force between charged particles.  The main difference is a minus sign, which makes like masses attract, while like charges repel.  In chemistry, people spend a lot of time thinking about charged particles obeying Schr\"odinger's equation, attracting or repelling each other electrostatically.  This approximation neglects magnetic fields, spin, and indeed anything related to the finiteness of the speed of light, but it's good enough explain quite a bit about atoms and molecules.

In this approximation, a collection of charged particles is described by a 
wavefunction $\psi$, which is a complex-valued function of all the particles' positions 
and also of time.   The basic idea is that $\psi$ obeys Schr\"odinger's
equation
\[ \frac{d \psi}{dt} = - i H \psi \]
where $H$ is an operator called the Hamiltonian, and I'm working in units where $\hbar = 1$.  

Does this equation succeeding in predicting $\psi$ at a later time given $\psi$ at time zero?  To answer this, we must first decide what kind of function $\psi$ should be, what concept of derivative applies to such funtions, and so on.  These issues were worked out by von Neumann and others starting in the late 1920s.    It required a lot of new mathematics.   Skimming the surface, we can say this.

At any time, we want $\psi$ to lie in the Hilbert space consisting of square-integrable functions of all the particle's positions.   We can then formally solve Schr\"odinger's equation as
$$  \psi(t) = \exp(-i t H) \psi(0) $$
where $\psi(t)$ is the solution at time $t$.  But for this to really work, we need $H$ to be a self-adjoint operator on the chosen Hilbert space.  The correct definition of `self-adjoint' is a bit subtler than what most physicists learn in a first course on quantum mechanics.  In particular, an operator can be superficially self-adjoint---the actual term for this is `symmetric'---but not truly self-adjoint.  

In 1951, based on earlier work of Rellich, Kato proved that $H$ is indeed self-adjoint for a collection of nonrelativistic quantum particles interacting via inverse-square forces \cite{Kato,Rellich}.   So, this simple model of chemistry works fine.  We can also conclude that 'celestial quantum mechanics' would dodge the nasty problems involving collisions or noncollision singularities that we saw in Newtonian gravity.   The reason, simply put, is the uncertainty principle.  

In the classical case, bad things happen because the energy is not bounded below.  A pair of classical particles attracting each other with an inverse square force law can have arbitrarily large \emph{negative} energy, simply by being very close to each other.   Noncollision singularities exploit this fact.  Since energy is conserved, if you have a way to make some particles get an arbitrarily large \emph{negative} energy, you can balance the books by letting others get an arbitrarily large \emph{positive} energy and shoot to infinity in a finite amount of time!

When we switch to quantum mechanics, the energy of any collection of particles becomes bounded below.  The reason is that to make the potential energy of two particles large and negative, they must be very close.  Thus, their difference in position must be very small.  In particular, this difference must be accurately known!  Thus, by the uncertainty principle, their difference in momentum must be very poorly known: at least one of its components must have a large standard deviation.   This in turn means that the expected value of the kinetic energy must be large.

This must all be made quantitative, to prove that as particles get close, the uncertainty principle provides enough positive kinetic energy to counterbalance the negative potential energy.  The Kato--Lax--Milgram--Nelson theorem \cite{ReedSimon}, a refinement of the original Kato--Rellich theorem, is the key to understanding this issue.  The Hamiltonian $H$ for a collection of particles interacting by inverse square forces can be written as $K + V$, where $K$ is an operator for the kinetic energy and $V$ is an operator for the potential energy.   With some clever work one can prove that for any $\epsilon > 0$, there exists $c > 0$ such that if $\psi$ is a smooth normalized wavefunction that vanishes at infinity and at points where particles collide, then
\[     | \langle \psi , V  \psi \rangle | \le \epsilon \langle \psi, K\psi \rangle + c. \]
Remember that $\langle \psi , V  \psi \rangle$ is the expected value of the potential energy, while $\langle \psi, K \psi \rangle$ is the expected value of the kinetic energy.  Thus, this inequality is a precise way of saying how kinetic energy triumphs over potential energy.

By taking $\epsilon = 1$, it follows that the Hamiltonian is bounded below on such 
states $\psi$:
\[     \langle \psi , H \psi \rangle \ge -c . \]
But the fact that the inequality holds even for smaller values of $\epsilon$ is the key to showing $H$ is `essentially self-adjoint'.  This means that while $H$ is not self-adjoint when defined only on smooth wavefunctions that vanish outside a bounded set and at points where particles collide, it has a unique self-adjoint extension to some larger domain.  Thus, we can unambiguously take this extension to be the true Hamiltonian for this problem.

To fully appreciate this, one needs to see what could have gone wrong.  Suppose space had an extra dimension.  In 3-dimensional space, Newtonian gravity obeys an inverse square force law because the area of a sphere is proportional to its radius squared.  In 4-dimensional space, the force obeys an inverse \emph{cube} law:
\[  F = -\frac{Gm_1 m_2}{r^3}  . \]
Using a cube instead of a square here makes the force stronger at short distances, with dramatic effects.  For example, even for the classical 2-body problem, the equations of motion no longer `almost always' have a well-defined solution for all times.  For an open set of initial conditions, the particles spiral into each other in a finite amount of time!   

The quantum version of this theory is also problematic.  The uncertainty principle is not enough to save the day.  The inequalities above no longer hold: kinetic energy does not triumph over potential energy.   The Hamiltonian is no longer essentially self-adjoint on the space of wavefunctions I described.  It has, in fact, \emph{infinitely many} self-adjoint extensions!  Each one describes \emph{different physics}: namely, a different choice of what happens when particles collide \cite{GTV,Gopalkrishnan}.  Moreover, when $G$ exceeds a certain critical value, the energy is no longer bounded below.  

The same problems afflict quantum particles interacting by the electrostatic force in 4d space, as long as some of the particles have opposite charges.   So, chemistry would be quite problematic in a world with four dimensions of space.  

With more dimensions of space, the situation becomes even worse.  This is part of a general pattern in mathematical physics: our struggles with the continuum tend to become worse in higher dimensions.  String theory and M-theory may provide exceptions.

\section{Classical electrodynamics of point particles}
\label{electrodynamics}

Now let us consider special relativity.  Special relativity prohibits instantaneous action at a distance.  Thus, most physicists believe that special relativity requires that forces be carried by fields, with disturbances in these fields propagating no faster than the speed of light.  The argument for this is not watertight, but we see to actually \emph{see} charged particles transmitting forces via a field, the electromagnetic field---that is, light.  So, most work on relativistic interactions brings in fields.

Classically, charged point particles interacting with the electromagnetic field are described by two sets of equations: Maxwell's equations and the Lorentz force law.  The first are a set of differential equations involving:
\begin{itemize}
\item the electric field $\vec E$ and mangetic field $\vec B$ (bundled together into the electromagnetic field $F$), and
\item the electric charge density $\rho$ and current density $\vec \jmath$ (bundled into another field called the `four-current' $J$). 
\end{itemize}
By themselves, these equations are not enough to completely determine the future given initial conditions.  In fact, you can choose $\rho$ and $\vec \jmath$ freely, subject
to the conservation law
\[   \frac{\partial \rho}{\partial t} + \nabla \cdot \vec \jmath = 0. \]
For any such choice, there exists a solution of Maxwell's equations for $t \ge 0$ given initial values for $\vec E$ and $\vec B$ that obey these equations at $t = 0$.  

Thus, to determine the future given initial conditions, we also need equations that say what $\rho$ and $\vec{\jmath}$ will do.  For a collection of charged point particles, they are determined by the curves in spacetime traced out by these particles.  The Lorentz force law says that the force on a particle of charge $e$ is
\[           \vec{F} = e (\vec{E} + \vec{v} \times \vec{B})  \]
where $\vec v$ is the particle's velocity and $\vec{E}$ and $\vec{B}$ are evaluated
at the particle's location.  From this law we can compute the particle's acceleration if we know its mass.   

The trouble starts when we try to combine Maxwell's equations and the Lorentz force law in a consistent way, with the goal being to predict the future behavior of the $\vec{E}$ and $\vec{B}$ fields, together with particles' positions and velocities, given all these quantities at $t = 0$.  Attempts to do this began in the late 1800s.  The drama continues today, with no definitive resolution!   Good accounts have been written by Feynman \cite{Feynman}, Pais \cite{Pais}, Janssen and Mecklenburg \cite{JM}, and Rohrlich \cite{Rohrlich1997}.  Here we can only skim the surface.

The first sign of a difficulty is this: the charge density and current associated to a charged particle are singular, vanishing off the curve it traces out in spacetime but `infinite' on this curve.  For example, a charged particle at rest at the origin has
\[        \rho(t,\vec x) = e \delta(\vec{x}), \qquad \vec{\jmath}(t,\vec{x}) = \vec{0} \]
where $\delta$ is the Dirac delta and $e$ is the particle's charge.   
This in turn forces the electric field to be singular at the origin.  The simplest solution of Maxwell's equations consistent with this choice of $\rho$ and $\vec\jmath$ is
\[  \vec{E}(t,\vec x) = \frac{e \hat{r}}{4 \pi \epsilon_0 r^2}, \qquad 
     \vec{B}(t,\vec x) = 0 \]
where $\hat{r}$ is a unit vector pointing away from the origin and $\epsilon_0$ is a constant called the permittivity of free space.  

In short, the electric field is `infinite', or undefined, at the particle's location.  So, it is unclear how to define the `self-force' exerted by the particle's own electric field on itself.
The formula for the electric field produced by a static point charge is really just our old friend, the inverse square law.  Since we had previously ignored the force of a particle 
on itself, we might try to continue this tactic now.   However, other problems intrude.

In relativistic electrodynamics, the electric field has energy density equal to 
\[         \frac{\epsilon_0}{2} |\vec{E}|^2 .   \]
Thus, the total energy of the electric field of a point charge at rest is proportional to
\[  \displaystyle{ \frac{\epsilon_0}{2} \int_{\mathbb{R}^3} |\vec{E}|^2 \, d^3 x =
\frac{e^2}{8 \pi \epsilon_0} \int_0^\infty \frac{1}{r^4} \, r^2 dr. } \]
But this integral diverges near $r = 0$, so the electric field of a charged particle 
has an infinite energy!   

How, if at all, does this cause trouble when we try to unify Maxwell's equations and the Lorentz force law?  It helps to step back in history.  In 1902, the physicist Abraham assumed that instead of a point, an electron is a sphere of radius $R$ with charge evenly distributed on its surface \cite{Abraham1902}. Then the energy of its electric field becomes finite, namely:
\[  E = \frac{e^2}{8 \pi \epsilon_0} \int_{R}^\infty \frac{1}{r^4} \, r^2 dr 
= \frac{1}{2} \frac{e^2}{4 \pi \epsilon_0 R} \]
where $e$ is the electron's charge. 

Abraham also computed the extra momentum that a \emph{moving} electron of this sort acquires due to its electromagnetic field.   He got it wrong because he didn't understand Lorentz transformations.  In 1904 Lorentz did the calculation right \cite{Lorentz}.
Using the relationship between velocity, momentum and mass, we can derive from his result a formula for the `electromagnetic mass' of the electron:
\[    m =  \frac{2}{3} \frac{e^2}{4 \pi \epsilon_0 R c^2}  \]
where $c$ is the speed of light.  We can think of this as the extra mass an electron acquires by carrying an electromagnetic field along with it.  

Putting the last two equations together, these physicists obtained a remarkable result:
\[  E = \frac{3}{4} mc^2 .\]
Then, in 1905, a fellow named Einstein came along and made it clear that the only reasonable relation between energy and mass is
\[   E = mc^2 .\]
What had gone wrong?  

In 1906, Poincar\'e figured out the problem \cite{Poincare}.  It is not a computational mistake, nor a failure to properly take special relativity into account.   The problem is that like charges repel, so if the electron were a sphere of charge it would explode without something to hold it together.   And that something---whatever it is---might have energy.  But their calculation ignored that extra energy.  

In short, the picture of the electron as a tiny sphere of charge, with nothing holding it together, is incomplete.   And the calculation showing $E = \frac{3}{4}mc^2$, together with special relativity saying $E = mc^2$, shows that this incomplete picture is actually inconsistent.  At the time, some physicists hoped that \emph{all} the mass of the electron could be accounted for by the electromagnetic field.  Their hopes were killed by
this discrepancy.

Nonetheless it is interesting to take the energy $E$ computed above, set it equal to $m_e c^2$ where $m_e$ is the electron's observed mass, and solve for the radius $R$.  The answer is 
\[     \displaystyle{  R = \frac{1}{8 \pi \epsilon_0} \frac{e^2}{m_e c^2} } \approx
1.4 \times 10^{-15} \; \mathrm{meters} .\]
In the early 1900s, this would have been a remarkably tiny distance: $0.00003$ times the Bohr radius of a hydrogen atom.  By now we know this is roughly the radius of a proton.  
We know the electron is not a sphere of this size.  So at present it makes more sense to treat the calculations so far as a prelude to some kind of limiting process where we take $R \to 0$.  These calculations teach us two lessons.  

First, the electromagnetic field energy approaches $+\infty$ as we let $R \to 0$, so it is challenging to take this limit and get a well-behaved physical theory.  One approach is to give a charged particle its own `bare mass' $m_\bare$ in addition to the mass $m_\elec$ arising from electromagnetic field energy, in a way that depends on $R$.  Then as we take the $R \to 0$ limit we can let $m_\bare \to -\infty$ in such a way that $m_\bare + m_\elec$ approaches a chosen limit $m$, the physical mass of the point particle.  This is an example of `renormalization'.

Second, it is wise to include conservation of energy-momentum as a requirement in addition to Maxwell's equations and the Lorentz force law.   Here is a more sophisticated way to phrase Poincar\'e's realization.  From the electromagnetic field one can compute a `stress-energy tensor' $T$, which describes the flow of energy and momentum through spacetime.   You can compute the total energy and momentum of the electromagnetic field by integrating $T$ over the hypersurface $t = 0$.  The resulting 4-vector will transform correctly under Lorentz transformations if the stress-energy tensor has vanishing divergence:
\[ \partial^\mu T_{\mu \nu} = 0 , \] 
where as usual we sum over repeated indices.  This equation says that energy and momentum are locally conserved.  However, this equation \emph{fails to hold} for a spherical shell of charge with no extra forces holding it together.  The reason is that in absence of extra forces, it violates conservation of momentum for charges to feel an electromagnetic force yet not accelerate.

So far we have only discussed the simplest situation: a single charged particle at rest, or moving at a constant velocity.  To go further, we can try to compute the acceleration of a small charged sphere in an arbitrary electromagnetic field.  Then, by taking the limit as the radius $r$ of the sphere goes to zero, perhaps we can obtain the law of motion for a charged point particle.   

In fact this whole program is fraught with difficulties, but physicists boldly go where
mathematicians fear to tread, and in a rough way this program was carried out already by Abraham \cite{Abraham1905} in 1905.  His treatment of special relativistic effects was wrong, but these were easily corrected; the real difficulties lie elsewhere.  In 1938 his calculations were carried out much more carefully---though still not rigorously---by Dirac \cite{Dirac}.  The resulting law of motion is thus called the `Abraham--Lorentz--Dirac force law'.

There are three key ways in which this law differs from our earlier naive statement of the Lorentz force law: 
\begin{itemize}
\item 
We must decompose the electromagnetic field in two parts, the `external' electromagnetic field $F_\ext$ and the field produced by the particle:
\[             F = F_\ext + F_\ret .  \]
Here $F_\ext$ is a solution of Maxwell's equations with $J = 0$, while $F_\ret$ is computed by convolving the particle's 4-current $J$ with a function called the `retarded Green's function'.  This breaks the time-reversal symmetry of the formalism, ensuring that radiation emitted by the particle moves outward as time goes into the future, not the past.   We then decree that the particle only feels a Lorentz force due to $F_\ext$, not $F_\ret$.  This avoids the problem that $F_\ret$ becomes infinite along the particle's path as $r \to 0$.
\item
Maxwell's equations say that an accelerating charged particle emits radiation, which carries energy-momentum.   Conservation of energy-momentum implies that there is a compensating force on the charged particle.  This is called the `radiation reaction'.  
So, in addition to the Lorentz force, there is a radiation reaction force.
\item 
As we take the limit $r \to 0$, we must adjust the particle's bare mass $m_\bare$ in 
such a way that its physical mass $m = m_\bare + m_\elec$ is held constant.   This involves letting $m_\bare \to -\infty$ as $m_\elec \to +\infty$.
\end{itemize}

It is easiest to describe the Abraham--Lorentz--Dirac force law using standard relativistic notation.  So, we switch to units where $c = 4 \pi \epsilon_0 = 1$, let $x^\mu$ denote the spacetime coordinates of a point particle, and use a dot to denote the derivative with respect to proper time.  Then the Abraham--Lorentz--Dirac force law says
\[   m \ddot{x}^\mu = e F_{\mathrm{ext}}^{\mu \nu} \,\dot{x}_\nu \; - \; 
\frac{2}{3}e^2 \ddot{x}^\alpha \ddot{x}_\alpha  \, \dot{x}^\mu \; + \; 
\frac{2}{3}e^2 \dddot{x}^\mu .\]
The first term at right is the Lorentz force, which looks more elegant in this new notation.  The second term acts to reduce the particle's velocity at a rate proportional to its velocity (as one would expect from friction), but also proportional to the squared magnitude of its acceleration.  This is the `radiation reaction'.

The last term, called the `Schott term', is the most shocking.  Unlike all familiar forces in classical mechanics, this involves the \emph{third} derivative of the particle's position!  This seems to shatter our original hope of predicting the electromagnetic field and the particle's position and velocity given their initial values.   Now it seems we need to specify the particle's initial position, velocity \emph{and acceleration}.  

Furthermore, unlike Maxwell's equations and the original Lorentz force law, 
the Abraham--Lorentz--Dirac force law is not symmetric under time reversal.   
If we take a solution, and replace $t$ with $-t$, the result is not a solution.  
Like the force of friction, radiation reaction acts to make a particle lose energy as it moves into the future, not the past.  The reason is that our assumptions 
have explicitly broken time reversal symmetry: the splitting $F = F_\ext + F_\ret$ 
says that a charged accelerating particle radiates into the future, creating the field 
$F_\ret$, and is affected only by the remaining electromagnetic field $F_\ext$.

Worse, the Abraham--Lorentz--Dirac force law has counterintuitive solutions.   Suppose for example that $F_\ext = 0$.   Besides the expected solutions where the particle's velocity is constant, there are solutions for which the particle accelerates
indefinitely, approaching the speed of light!   These are called `runaway solutions'.
In a runaway solution, the acceleration as measured in the frame of reference of the particle grows exponentially with the passage of proper time.  

So, the notion that special relativity might help us avoid the pathologies of
Newtonian point particles interacting gravitationally---solutions where 
particles shoot to infinity in finite time---is cruelly mocked by the 
Abraham--Lorentz--Dirac force law.  Particles cannot move faster than light, but even a \emph{single} particle can extract an arbitrary amount of energy-momentum from the electromagnetic field in its immediate vicinity and use this to propel itself forward at speeds approaching that of light.  

The energy stored in the field near the particle is sometimes called `Schott energy'.    The Schott term describes how this energy can be converted into kinetic energy for the particle.  The details have been nicely explained by Gr\o n \cite{Gron}.

Even worse, suppose we generalize the framework to include more than one
particle.  Arguments for the Abraham--Lorentz--Dirac force law can be generalized to  this case, and the result is simply that each particle obeys this law with an 
external field $F_\ext$ that includes the fields produced by all the other particles.  
But a problem appears when we use this law to compute the motion of two particles of opposite charge.   To simplify the calculation, suppose they are located symmetrically with respect to the origin, with equal and opposite velocities and accelerations.   Suppose the external field felt by each particle is solely the field created by the other particle.   Since the particles have opposite charges, they should attract each other.  However, one can prove they will never collide.  In fact, if at any time they are moving toward each other, they will later \emph{turn around and move away from each other at ever-increasing speed!}

This fact was discovered by Eliezer \cite{Eliezer} in 1943.  It is so counterintuitive that several proofs were required before physicists believed it.  A self-contained proof and review of the literature can be found in Parrott's book \cite{Parrott}, along with a discussion of the runaway solutions mentioned earlier.

None of these strange phenomena have ever been seen experimentally.  Faced with this
problem, physicists have naturally looked for ways out.   First, why not simply \emph{cross out} the Schott term in the Abraham--Lorentz--Dirac 
force?  Unfortunately the resulting simplified equation
\[    m \ddot{x}^\mu = e F_{\ext}^{\mu \nu} \, \dot{x}_\nu   \;- \;
\frac{2}{3}e^2 \ddot{x}^\alpha \ddot{x}_\alpha \, \dot{x}^\mu \]
has only trivial solutions.  The reason is that with the particle's path parametrized by 
proper time, the vector $\dot{x}^\mu$ has constant length, so the vector $\ddot{x}^\mu$ is orthogonal to $\dot{x}^\mu$ .  So is the vector 
$F_{\ext}^{\mu \nu} \, \dot{x}_\nu$, because $F_{\ext}$ is an antisymmetric
tensor.  So, the last term must be zero, which implies $\ddot{x} = 0$, 
which in turn implies that all three terms must vanish.
 
Another possibility is that some assumption made in deriving the Abraham--Lorentz--Dirac force law is incorrect.  Of course the theory is \emph{physically} incorrect, in that it ignores quantum mechanics, but that is not the issue.  The issue here is one of mathematical physics, of trying to formulate a well-behaved classical theory that describes charged point particles interacting with the electromagnetic field.  If we can prove this is impossible, we will have learned something.  But perhaps there is a loophole. The original arguments for the Abraham--Lorentz--Dirac force law are by no means mathematically rigorous.  They involve a delicate limiting procedure, and 
approximations that were believed, but not proved, to become perfectly accurate 
in the $r \to 0$ limit.  Could these arguments conceal a mistake?  

Calculations involving a spherical shell of charge has been improved by a series of 
authors, and are nicely summarized by Rohrlich \cite{Rohrlich1997,Rohrlich1999}. 
In all these calculations, nonlinear powers of the acceleration and its time derivatives
are neglected, and one hopes this is acceptable in the $r \to 0$ limit.   

Dirac \cite{Dirac}, struggling with renormalization in quantum field theory, took a
different tack.  Instead of considering a sphere of charge, he treated the 
electron as a point from the very start.  However, he studied the flow of energy-momentum across the surface of a tube of radius $r$ centered on the electron's path.
By computing this flow in the limit $r \to 0$, and using conservation of energy-momentum, he attempted to derive the force on the electron.  He did not obtain a 
unique result, but the simplest choice gives the Abraham--Lorentz--Dirac equation.  
More complicated choices typically involve nonlinear powers of the acceleration and its time derivatives.

Since this work, many authors have tried to simplify Dirac's rather complicated calculations 
and clarify his assumptions.  Parrott's book is a good guide to much of this work 
\cite{Parrott}.  But more recently, Kijowski and coauthors have made impressive progress in a series of papers that solve many of the problems we have seen \cite{GKZ1998,Kijowski1994a,Kijowski1994b,KP2003,KP2007}.

Kijowski's key idea is to impose conditions on precisely how the electromagnetic
field is allowed to behave near the path traced out by a charged point particle.  He decomposes the field into a `regular' part and a `singular' part:
\[            F = F_\reg + F_\sing . \]
Here $F_\reg$ is smooth everywhere, while $F_\sing$ is singular near the particle's
path, but only in a carefully prescribed way.   Roughly, at each moment, in the particle's instantaneous rest frame, the singular part of its electric field consists of the familiar
term proportional to $1/r^2$, together with a term proportional to $1/r^3$ which depends on the particle's acceleration.  No other singularities are allowed. 

On the one hand, this eliminates the ambiguities mentioned earlier: in the end, there are no `nonlinear powers of the acceleration and its time derivatives' in Kijowski's force law.  On the other hand, this avoids breaking time reversal symmetry, as the earlier splitting $F = F_\ext + F_\ret$ did.  

Next, Kijowski defines the energy-momentum of a point particle to be $m \dot{x}$,
where $m$ is its physical mass.   He defines the energy-momentum of the electromagnetic field to be just that due to $F_\reg$, not $F_\sing$.   This amounts to eliminating the infinite `electromagnetic mass' of the charged particle. He then shows that Maxwell's equations and conservation of total energy-momentum imply an equation of motion for the particle.

This equation is very simple:
\[    m \ddot{x}^\mu = e F_{\reg}^{\mu \nu} \,\dot{x}_\nu  . \]
It is just the Lorentz force law.    Since the troubling Schott term is gone,
this is a second-order differential equation.  Thus, we can hope that to predict the future behavior of the electromagnetic field, together with the particle's position and velocity, given all these quantities at $t = 0$.  

And indeed this is true!  In 1998, together with Gittel and Zeidler, Kijowski proved that initial data of this sort, obeying the careful restrictions on allowed singularities of the electromagnetic field, determine a unique solution of Maxwell's equations and the Lorentz force law, at least for a short amount of time \cite{GKZ1998}.   Even better, all this remains true for any number of particles.

There are some obvious questions to ask about this new approach.  In the Abraham--Lorentz--Dirac force law, the acceleration was an independent variable that needed to be specified at $t = 0$ along with position and momentum.   This problem disappears in Kijowski's approach.  But how?  I mentioned that the singular part of the electromagnetic field, $F_\sing$, depends on the particle's acceleration.  But more is true: the particle's acceleration is completely determined by $F_\sing$.  So, the particle's acceleration is not an independent variable because it is \emph{encoded into the electromagnetic field}.

Another question is: where did the radiation reaction go?  The answer is: we can see it if we go back and decompose the electromagnetic field as $F_\ext + F_\ret$ as we had before.   If we take the law
\[    m \ddot{x}^\mu = e F_{\reg}^{\mu \nu} \dot{x}_\nu  \]
and rewrite it in terms of $F_\ext$, we recover the original Abraham--Lorentz--Dirac
law, including the radiation reaction term and Schott term.

Unfortunately, this means that `pathological' solutions where particles extract arbitrary amounts of energy from the electromagnetic field are still possible.   A related problem
is that apparently nobody has yet proved solutions exist for all time.  Perhaps a singularity worse than the allowed kind could develop in a finite amount of time---for example, when particles collide.  

Thus, classical point particles interacting with the electromagnetic field still present serious challenges to the physicist and mathematician. When you have an infinitely small charged particle right next to its own infinitely strong electromagnetic field, trouble can break out very easily!

Finally, I should also mention attempts, working within the framework of special relativity, to get rid of fields and have particles interact with each other directly.  For example, in 1903 Schwarzschild \cite{Schwarzschild} introduced a framework in which charged particles exert an electromagnetic force on each other, with no mention of fields.  In this setup, forces are transmitted not instantaneously but at the speed of light: the force on one particle at spacetime point $x$ depends on the motion of some other particle at spacetime point $y$ only if the vector $x - y$ is lightlike.  Later Fokker and Tetrode \cite{Fokker,Tetrode} derived this force law from a principle of least action.  In 1949, Feynman and Wheeler checked that this formalism gives results compatible with the usual approach to electromagnetism using fields, except for several points:
\begin{itemize}
\item Each particle exerts forces only on \emph{other} particles, so we avoid the thorny issue of how a point particle responds to the electromagnetic field produced by itself.
\item There are no electromagnetic fields not produced by particles: for example, the theory does not describe the motion of a charged particle in an `external electromagnetic
field'.
\item The principle of least action guarantees that `if $A$ affects $B$ then $B$ affects
$A$'.  So, if a particle at $x$ exerts a force on a particle at a point $y$ in its future lightcone, the particle at $y$ exerts a force on the particle at $x$ in its past lightcone.  This raises the issue of `reverse causality', which Feynman and Wheeler address.
\end{itemize}
Besides the reverse causality issue, perhaps one reason this approach has 
not been more pursued is that it does not admit a Hamiltonian formulation in terms
of particle positions and momenta.  Indeed, there are a number of `no-go theorems' for relativistic multiparticle Hamiltonians \cite{CJS,Leutwyler}, saying that these can only 
describe noninteracting particles.   So, most work that takes \emph{both} quantum mechanics and special relativity into account uses fields.  Indeed, in quantum electrodynamics, even the charged point particles are replaced by fields.

\section{Quantum field theory}
\label{qft}

When we study charged particles interacting electromagnetically in a way that
takes both quantum mechanics and special relativity into account, we are led to
quantum field theory.  The ensuing problems are vastly more complicated than in any of the physical theories discussed so far.  They are also more consequential, since at present quantum field theory is our best description of all known forces except gravity.  As a result, many of the best minds in 20th-century mathematics and physics have joined the fray, and it is impossible here to give more than a quick summary of the situation. This is especially true because the final outcome of the struggle is not yet known.

It is ironic that quantum field theory originally emerged as a \emph{solution} to a problem involving the continuum nature of spacetime, now called the `ultraviolet catastrophe'.  In classical electromagnetism, a box with mirrored walls containing only radiation acts like a collection of harmonic oscillators, one for each vibrational mode of the electromagnetic field.  If we assume waves can have arbitrarily small wavelengths, there are infinitely many of these oscillators.  In classical thermodynamics, a collection of harmonic oscillators in thermal equilibrium will share the available energy equally: this result is called the `equipartition theorem'.  

Taken together, these principles lead to a dilemma worthy of Zeno.  The energy in the box must be divided into an infinite number of equal parts.  If the energy in each part is nonzero, the total energy in the box must be infinite.  If it is zero, there can be no energy in the box.  

For the logician, there is an easy way out: perhaps a box of electromagnetic radiation can only be in thermal equilibrium if it contains no energy at all!   But this solution cannot satisfy the physicist, since it does not match what is actually observed.  In reality, any nonnegative amount of energy is allowed in thermal equilibrium.  

Experiment also rules out another cheap solution: simply forbidding, by fiat, waves with wavelength shorter than some fixed length.  This makes the infinities go away.  However, we find that for any nonzero temperature, most of the radiation in a mirrored box will have very short wavelength.  This is not what is observed.

The right way out of the dilemma was to change our concept of the harmonic oscillator.  Planck did this in 1900, almost without noticing it \cite{Planck}.  Classically, a harmonic oscillator can have any nonnegative amount of energy.  Planck instead treated the energy 

\begin{quote}
... not as a continuous, infinitely divisible quantity, but as a discrete quantity composed of an integral number of finite equal parts.
\end{quote}

In modern notation, the allowed energies of a quantum harmonic oscillator are integer multiples of $\hbar \omega$, where $\omega$ is the oscillator's frequency and $\hbar$ is a new constant of nature, named after Planck.   When energy can only take such discrete values, the equipartition theorem no longer applies.  Instead, the principles of thermodynamics imply that there is a well-defined thermal equilibrium in which vibrational modes with shorter and shorter wavelengths, and thus higher and higher energies, hold less and less of the available energy.   The results agree with experiments when the constant $\hbar$ is given the right value.

The full import of what Planck had done became clear only later, starting with Einstein's 1905 paper on the photoelectric effect \cite{Einstein}.  Here he proposed that the discrete energy steps actually arise because light comes in particles, now called `photons', with a photon of frequency $\omega$ carrying energy $\hbar \omega$.   It was even later that Ehrenfest emphasized the role of the equipartition theorem in the original dilemma, and called this dilemma the `ultraviolet catastrophe'.  As usual, the actual history is more complicated than the textbook summaries \cite{Kragh}.

The theory of the `free' quantum electromagnetic field---that is, photons not interacting with charged particles---is now well understood.   It is a bit tricky to deal with an infinite collection of quantum harmonic oscillators, but since each evolves independently from all the rest, the issues are manageable.  Many advances in analysis were required to tackle these issues in a rigorous way, but they were erected on a sturdy scaffolding of algebra.  The reason is that the quantum harmonic oscillator is exactly solvable in terms of well-understood functions, and so is the free quantum electromagnetic field.   By the 1930s,  physicists knew precise formulas for the answers to more or less any problem involving the free quantum electromagnetic field.  The challenge to mathematicians was then to find a coherent mathematical framework that takes us to these answers starting from clear assumptions.  This challenge was taken up and completely met by the mid-1960s \cite{BSZ}.

However, for physicists, the free quantum electromagnetic field is just the starting-point, since this field obeys a quantum version of Maxwell's equations where the charge density and current density vanish.   Far more interesting is `quantum electrodynamics', or QED, where we also include fields describing charged particles---for example, electrons and their antiparticles,  positrons---and try to impose a quantum version of the full-fledged Maxwell equations. Nobody has found a fully rigorous formulation of QED, nor has anyone proved such a thing cannot be found.

QED is part of a more complicated quantum field theory, the Standard Model, which describes the electromagnetic, weak and strong forces, quarks and leptons, and the Higgs boson.  It is widely regarded as our best theory of elementary particles.  Unfortunately, nobody has found a rigorous formulation of this theory either, despite decades of hard work by many smart physicists and mathematicians.

To spur progress, the Clay Mathematics Institute has offered a million-dollar prize for anyone who can prove a widely believed claim about a class of quantum field theories called `pure Yang--Mills theories' \cite{ClayYM}.  A good example is the fragment of the Standard Model that describes only the strong force---or in other words, only gluons.  Unlike photons in QED, gluons interact with each other.  To win the prize, one must prove that the theory describing them is mathematically consistent and that it describes a world where the lightest particle is a `glueball': a blob made of gluons, with mass strictly greater than zero.  This theory is considerably simpler than the Standard Model.  However, it is already very challenging.

This is not the only million-dollar prize that the Clay Mathematics Institute is offering for struggles with the continuum.  They are also offering one for a proof of global existence of solutions to the Navier--Stokes equations for fluid flow \cite{ClayNS}.  However, their quantum field theory challenge is the only one for which the problem statement is not completely precise.  The Navier--Stokes equations are a collection of partial differential equations for the velocity and pressure of a fluid.  We know how to precisely phrase the question of whether these equations have a well-defined solution for all time given smooth initial data.  Describing a quantum field theory is a trickier business!

To be sure, there are a number of axiomatic frameworks for quantum field theory \cite{Haag,StreaterWightman}.  We can prove physically interesting theorems  from these axioms, and also rigorously construct some quantum field theories obeying these axioms \cite{BSZ,GlimmJaffe,Rivasseau}.  The easiest are the free theories,  which describe noninteracting particles.  There are also examples of rigorously  construted quantum field theories that describe interacting particles in fewer than 4 spacetime dimensions.  However, no quantum field theory that describes interacting particles in 4-dimensional spacetime has been proved to obey the usual axioms.   Thus, much of the wisdom of physicists concerning quantum field theory has not been fully transformed into rigorous mathematics.  

Worse, the question of whether a particular quantum field theory studied by physicists obeys the usual axioms is not completely precise---at least, not yet.  The problem is that going from the physicists' formulation to a mathematical structure that might or might not obey the axioms involves some choices.

This is not a cause for despair; it simply means that there is much work left to be done. In practice, quantum field theory is marvelously good for calculating answers to many physics questions.  The answers involve approximations.   These approximations seem to work very well: that is, they answers that match experiments.  Unfortunately we do not fully understand, in a mathematically rigorous way, what these approximations are supposed to be approximating.

How could this be?  I will try to sketch some of the key issues in the case of quantum electrodynamics.  The history of QED has been nicely told by Schweber \cite{Schweber}, so I will focus on concepts rather than the history, and hope that experts forgive me for cutting corners and trying to get across the basic ideas at the expense of many technical details.  The nonexpert is encouraged to fill in the gaps with the help of some textbooks, for example those of Zee \cite{Zee}, Peskin and Schroeder \cite{PS}, Itzykson and Zuber \cite{IZ}, or for a more mathematical view, Ticciati \cite{Ticciati}.

QED involves just one dimensionless parameter, the fine structure constant:
\[   \alpha = \frac{1}{4 \pi \epsilon_0} \frac{e^2}{\hbar c} \approx \frac{1}{137.036} .\]
We can think of $\alpha^{1/2}$ as a dimensionless version of the electron charge.  It  says how strongly electrons and photons interact.   

Nobody knows why the fine structure constant has the value it does.  In computations, we are free to treat it as an adjustable parameter.  If we set it to zero, quantum electrodynamics reduces to a free theory, where photons and electrons do not interact with each other.  A standard strategy in QED is to take advantage of the fact that the fine structure constant is small and expand answers to physical questions as power series in $\alpha^{1/2}$.  This is called `perturbation theory', and it allows us to exploit our knowledge of free theories. 

One of the main questions we try to answer in QED is this: if we start with some particles with specified energy-momenta in the distant past, what is the probability that they will turn into certain other particles with certain other energy-momenta in the distant future?   As usual, we compute this probability by first computing a complex amplitude and then taking the square of its absolute value.  The amplitude, in turn, is computed as a power series in $\alpha^{1/2}$.  

The term of order $\alpha^{n/2}$ in this power series is a sum over Feynman diagrams with $n$ vertices.  For example, suppose we are computing the amplitude for two electrons with some specified energy-momenta to interact and become two electrons with some other energy-momenta.  One Feynman diagram appearing in the answer is this:
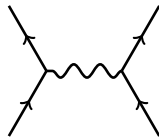
\[
\begin{tikzpicture}[line width=1 pt, scale=1]
	\draw[fermion] (-120:1)--(0,0);
	\draw[fermionbar] (120:1)--(0,0);
	\draw[vector] (0:1)--(0,0);
 \begin{scope}[shift={(1,0)}]
	\draw[fermion] (-60:1)--(0,0);
	\draw[fermionbar] (60:1)--(0,0);
\end{scope}
\end{tikzpicture}
\]
Here the electrons exhange a single photon.  Since this diagram has two vertices, it contributes a term of order $\alpha$.   The electrons could also exchange two photons:
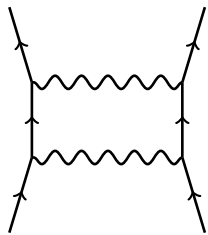
\[
\begin{tikzpicture}[line width=1 pt, scale=1]
		\draw[fermion] (2.3,-1) -- (2,0);
		\draw[fermion] (2,0) -- (2,1);
		\draw[fermion] (2,1) -- (2.3,2);
		\draw[fermion] (-.3,-1) -- (0,0);
		\draw[fermion] (0,0) -- (0,1);
		\draw[fermion] (0,1) -- (-.3,2);
		\draw[vector] (2,0) -- (0,0);
		\draw[vector] (2,1) -- (0,1);
	\end{tikzpicture}
\]
giving a term of $\alpha^2$.  A more interesting term of order $\alpha^2$ is this:
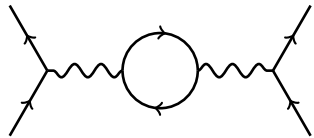
\[
\begin{tikzpicture}[line width=1 pt]
	\draw[fermion] (-120:1)--(0,0);
	\draw[fermionbar] (120:1)--(0,0);
	\draw[vector] (0:1)--(0,0);
	\draw[fermion] (1,0) arc (180:0:.5);
	\draw[fermion] (2,0) arc (0:-180:.5);
	\draw[vector] (2,0) --(3,0);
\begin{scope}[shift={(3,0)}]
	\draw[fermion] (-60:1)--(0,0);
	\draw[fermionbar] (60:1)--(0,0);
\end{scope}
\end{tikzpicture}
\]
Here the electrons exchange a photon that splits into an electron-positron pair and then recombines.  There are infinitely many diagrams with two electrons coming in and two going out.  However, there are only finitely many with $n$ vertices.  Each of these contributes a term proportional to $\alpha^{n/2}$ to the amplitude.

In general, the external edges of these diagrams correspond to the experimentally observed particles coming in and going out.  The internal edges correspond to `virtual particles': that is, particles that are not directly seen, but appear in intermediate steps of a process.

Each of these diagrams is actually a notation for an integral. There are systematic rules for writing down the integral starting from the Feynman diagram \cite{IZ,PS}.  To do this, we first label each edge of the Feynman diagram with an energy-momentum, a variable $p \in \R^4$.   The integrand, which we shall not describe here, is a function of all these energy-momenta.  In carrying out the integral, the energy-momenta of the external edges are held fixed, since these correspond to the experimentally observed particles coming in and going out.  We integrate over the energy-momenta of the internal edges, which correspond to virtual particles, while requiring that energy-momentum is conserved at each vertex.  

However, there is a problem: the integral typically diverges!  Whenever a Feynman diagram contains a loop, the energy-momenta of the virtual particles in this loop can be arbitrarily large.  Thus, we are integrating over an infinite region.  In principle the integral could still converge if the integrand goes to zero fast enough.   However, we rarely have such luck.

What does this mean, physically?  It means that if we allow virtual particles with arbitrarily large energy-momenta in intermediate steps of a process, there are `too many ways for this process to occur', so the amplitude for this process diverges.  

Ultimately, the continuum nature of spacetime is to blame.  In quantum mechanics, particles with large momenta are the same as waves with short wavelengths.   Allowing light with arbitrarily short wavelengths created the ultraviolet catastrophe in classical electromagnetism.  Quantum electromagnetism averted that catastrophe---but the problem returns in a different form as soon as we study the interaction of photons and charged particles.

Luckily, there is a strategy for tackling this problem. The integrals for Feynman diagrams become well defined if we impose a `cutoff', integrating only over energy-momenta $p$ in some bounded region, say a ball of some large radius $\Lambda$.   In quantum theory, a particle with momentum of magnitude greater than $\Lambda$ is the same as a wave with wavelength less than $\hbar/\Lambda$.    Thus, imposing the cutoff amounts to ignoring waves of short wavelength---and for the same reason, ignoring waves of high frequency.   We obtain well-defined answers to physical questions when we do this.  Unfortunately the answers depend on $\Lambda$, and if we let $\Lambda \to \infty$, they diverge.

However, this is not the correct limiting procedure.  Indeed, among the quantities that we can compute using Feynman diagrams are the charge and mass of the electron!   Its charge can be computed using diagrams in which an electron emits or absorbs a photon:
{\boldmath
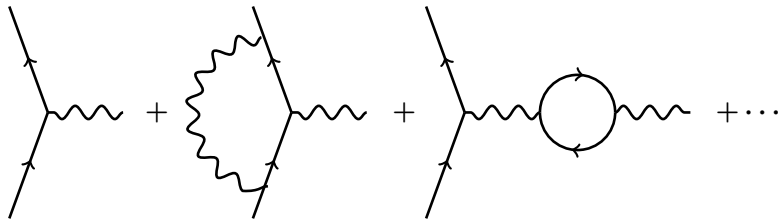
\[   \begin{tikzpicture}[line width=1 pt, scale=1]
	\draw[fermion] (-110:1.5)--(0,0);
	\draw[fermionbar] (110:1.5)--(0,0);
	\draw[vector] (0:1)--(0,0);
\node at (0:1.5) { $+ $\;};
\end{tikzpicture}
\begin{tikzpicture}[line width=1 pt, scale=1]
	\draw[fermion] (-110:1.5)--(0,0);
	\draw[fermionbar] (110:1.5)--(0,0);
      \draw[vector] (-0.4,1) arc (100:265:1);
	\draw[vector] (0:1)--(0,0);
\node at (0:1.5) { $+$ };
\end{tikzpicture}
 \begin{tikzpicture}[line width=1 pt, scale=1]
      \draw[fermion] (-110:1.5)--(0,0);
	\draw[fermionbar] (110:1.5)--(0,0);
	\draw[vector] (0:1)--(0,0);
	\draw[fermion] (1,0) arc (180:0:.5);
	\draw[fermion] (2,0) arc (0:-180:.5);
	\draw[vector] (2,0) --(3,0);
  	\node at (0:3.8) { $+ \cdots$};
\end{tikzpicture}
 \]
}
Similarly, its mass can be computed using a sum over Feynman diagrams where one electron comes in and one goes out.  

The interesting thing is this: to do these calculations, we must start by assuming some charge and mass for the electron---but the charge and mass we \emph{get out} of these calculations do not equal the masses and charges we \emph{put in!}   

The reason is that virtual particles affect the observed charge and mass of a particle.   Heuristically, at least, we should think of an electron as surrounded by a cloud of virtual particles.  These contribute to its mass and `shield' its electric field, reducing its observed charge.  It takes some work to translate between this heuristic story and actual Feynman diagram calculations, but it can be done.  

Thus, there are two different concepts of mass and charge for the electron.  The numbers we put into the QED calculations are called the `bare' charge and mass, $e_\bare$ and $m_\bare$.    Poetically speaking, these are the charge and mass we would see if we could strip the electron of its virtual particle cloud and see it in its naked splendor.   The numbers we get out of the QED calculations are called the `renormalized' charge and mass, $e_\ren$ and $m_\ren$.  These are computed by doing a sum over Feynman diagrams.  So, they take virtual particles into account.  These are the charge and mass of the electron clothed in its cloud of virtual particles.  It is these quantities, not the bare quantities, that should agree with experiment.  

Thus, the correct limiting procedure in QED calculations is a bit subtle.   For any value of $\Lambda$ and any choice of $e_\bare$ and $m_\bare$, we compute $e_\ren$ and $m_\ren$.  The necessary integrals all converge, thanks to the cutoff.   We choose $e_\bare$ and $m_\bare$ so that $e_\ren$ and $m_\ren$ agree with the experimentally observed charge and mass of the electron.  The bare charge and mass chosen this way depend on $\Lambda$, so call them $e_\bare(\Lambda)$ and $m_\bare(\Lambda)$.   

Next, suppose we want to compute the answer to some other physics problem using QED.  We do the calculation with a cutoff $\Lambda$, using $e_\bare(\Lambda)$ and $m_\bare(\Lambda)$ as the bare charge and mass in our calculation.  Then we take the limit $\Lambda \to \infty$.  

In short, rather than simply fixing the bare charge and mass and letting $\Lambda \to \infty$, we cleverly adjust the bare charge and mass as we take this limit.  This procedure is called `renormalization', and it has a complex and fascinating history \cite{Brown}.  There are many technically different ways to carry out renormalization, and our account so far neglects many important issues.   Let us mention three of the simplest.  
 
First, besides the classes of Feynman diagrams already mentioned, we must also consider those where one photon goes in and one photon goes out, such as this:
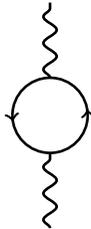
\[
\begin{tikzpicture}[line width=1 pt, scale= 1]
	\draw[vector] (1:0)--(0,1);
	\draw[fermion] (0,1) arc (-90:90:.5);
	\draw[fermion] (0,2) arc (90:270:.5);
	\draw[vector] (0,2) --(0,3);
\end{tikzpicture}
\]
These affect properties of the photon, such as its mass.  Since we want the photon to be massless in QED, we have to adjust parameters as we take $\Lambda \to \infty$ to make sure we obtain this result.  We must also consider Feynman diagrams where nothing comes in and nothing comes out---so-called `vacuum bubbles'---and make these behave correctly as well.

Second, the procedure just described, where we impose a `cutoff' and integrate over energy-momenta $p$ lying in a ball of radius $\Lambda$, is not invariant under Lorentz transformations.   Indeed, any theory featuring a smallest time or smallest distance violates the principles of special relativity: thanks to time dilation and Lorentz contractions, different observers will disagree about times and distances.  We could accept that Lorentz invariance is broken by the cutoff and hope that it is restored in the $\Lambda \to \infty$ limit, but physicists prefer to maintain symmetry at every step of the calculation.  This requires some new ideas: for example, replacing Minkowski spacetime with 4-dimensional Euclidean space.  In 4-dimensional Euclidean space, Lorentz transformations are replaced by rotations, and a ball of radius $\Lambda$ is a rotation-invariant concept.   To do their Feynman integrals in Euclidean space, physicists often let time take imaginary values.  They do their calculations in this context and then transfer the results back to Minkowski spacetime at the end.   Luckily, there are theorems justifying this procedure \cite{GlimmJaffe, Haag, StreaterWightman}.

Third, besides infinities that arise from waves with arbitrarily short wavelengths, there are infinities that arise from waves with arbitrarily \emph{long} wavelengths.   The former are called `ultraviolet divergences'.  The latter are called `infrared divergences', and they afflict theories with massless particles, like the photon.  For example, in QED the collision of two electrons will emit an infinite number of photons with very long wavelengths and low energies, called `soft photons'.  In practice this is not so bad, since any experiment can only detect photons with energies above some nonzero value.   However, infrared divergences are conceptually important.  It seems that in QED any electron is inextricably accompanied by a cloud of soft photons \cite{Buchholz}.   This is distinct from the `virtual particle cloud' that I mentioned before: these are real particles, emitted by the electron whenever it accelerates.

Now let us summarize what we do and do not know about perturbation theory in QED.  On the bright side, thanks to the efforts of many brilliant physicists and mathematicians, QED has been proved to be `perturbatively renormalizable' \cite{FHRW,Scharf}.  This means that we can indeed carry out the procedure roughly sketched above, obtaining answers to physical questions as power series in $\alpha^{1/2}$.  On the dark side, we do not know if these power series converge.   In fact, it is widely believed that they diverge!   This puts us in a curious situation.  

A good example is the magnetic dipole moment of the electron.  An electron, being a charged particle with spin, has a magnetic field.   A classical computation says that its magnetic dipole moment is
\[   \vec{\mu} = -\frac{e}{2m_e} \vec{S} \]
where $\vec{S}$ is its spin angular momentum.  Quantum effects correct this computation, giving
\[  \vec{\mu} = -g \frac{e}{2m_e} \vec{S} \]
for some constant $g$ called the `gyromagnetic ratio'.   This constant can be computed using QED as a sum over Feynman diagrams in which an electron exchanges a single photon with a massive charged particle:
{\boldmath
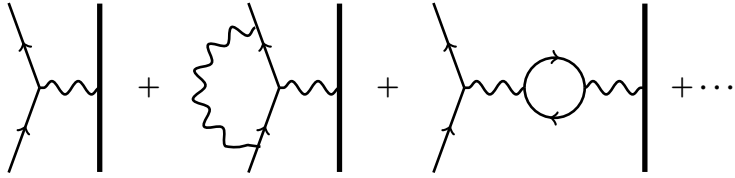
\[   \begin{tikzpicture}[line width=1 pt, scale=0.8]
	\draw[fermion] (-110:1.5)--(0,0);
	\draw[fermionbar] (110:1.5)--(0,0);
	\draw[vector] (0:1)--(0,0);
       \draw [line width = 2 pt] (1,-1.4) -- (1,1.4);
\node at (0:2) { $+ $\;\;\;};
\end{tikzpicture}
\begin{tikzpicture}[line width=1 pt, scale=0.8]
	\draw[fermion] (-110:1.5)--(0,0);
	\draw[fermionbar] (110:1.5)--(0,0);
      \draw[vector] (-0.4,1) arc (100:265:1);
	\draw[vector] (0:1)--(0,0);
      \draw [line width = 2 pt] (1,-1.4) -- (1,1.4);
\node at (0:2) { $+$ \;\; };
\end{tikzpicture}
 \begin{tikzpicture}[line width=1 pt, scale=0.8]
      \draw[fermion] (-110:1.5)--(0,0);
	\draw[fermionbar] (110:1.5)--(0,0);
	\draw[vector] (0:1)--(0,0);
	\draw[fermion] (1,0) arc (180:0:.5);
	\draw[fermion] (2,0) arc (0:-180:.5);
	\draw[vector] (2,0) --(3,0);
        \draw [line width = 2 pt] (3,-1.4) -- (3,1.4);
  	\node at (0:4) { $+ \cdots$};
\end{tikzpicture}
 \]
}
The answer is a power series in $\alpha^{1/2}$, but since all these diagrams have an even number of vertices, it only contains integral powers of $\alpha$.  The lowest-order term gives simply $g = 2$.  In 1948, Schwinger \cite{Schwinger} computed the next term and found a small correction to this simple result:
\[   g = 2 + \frac{\alpha}{\pi} \approx 2.00232 .\]
By now have people have computed $g$ up to order $\alpha^5$.  This requires computing over 13,000 integrals, one for each Feynman diagram of the above form with up to 10 vertices \cite{AHKN}.  The answer agrees very well with experiment: in fact, if we also take other Standard Model effects into account we get agreement to roughly one part in $10^{12}$.  This is the most accurate prediction in all of science!

However, if we continue adding up terms in this power series, there is no guarantee that the answer converges.  Indeed, in 1952 Freeman Dyson \cite{Dyson} gave a heuristic argument that makes physicists expect that the series \emph{diverges}, along with most other power series in QED.  

The argument goes as follows. If these power series converged for small positive $\alpha$, they would have a nonzero radius of convergence, so they would also converge for small negative $\alpha$.  Thus, QED would make sense for small negative values of $\alpha$, which correspond to \emph{imaginary} values of the electron's charge.  If the electron had an imaginary charge, electrons would attract each other electrostatically, since the usual repulsive force between them is proportional to $e^2$.  Thus, if the power series converged, we would have a theory like QED for electrons that attract rather than repel each other.

However, there is a good reason to believe that QED cannot make sense for electrons that attract.  The reason is that it describes a world where the vacuum is unstable. That is, there would be states with arbitrarily large negative energy containing many electrons and positrons.  Thus, we expect that the vacuum could spontaneously turn into electrons and positrons together with photons (to conserve energy).   Of course, this is not a rigorous proof that the power series in QED diverge: just an argument that it would be strange if they did not.

To see why electrons that attract could have arbitrarily large negative energy, consider a state $\psi$ with a large number $N$ of such electrons inside a ball of radius $R$.   We require that these electrons have small momenta, so that nonrelativistic quantum mechanics gives a good approximation to the situation.   Since its momentum is small, the kinetic energy of each electron is a small fraction of its rest energy $m_e c^2$.   If we let $\langle \psi, E \psi\rangle$ be the expected value of the total rest energy and kinetic energy of all the electrons, it follows that $\langle \psi, E\psi \rangle$ is approximately proportional to $N$.

The Pauli exclusion principle puts a limit on how many electrons with momentum below some bound can fit inside a ball of radius $R$.    This number is asymptotically proportional to the volume of the ball.  Thus, we can assume $N$ is approximately proportional to $R^3$.    It follows that $\langle \psi, E \psi \rangle$ is approximately proportional to $R^3$.   

There is also the negative potential energy to consider.   Let $V$ be the operator for potential energy.  Since we have $N$ electrons attracted by an $1/r$ potential, and each pair contributes to the potential energy, we see that $\langle \psi , V \psi \rangle$ is approximately proportional to $-N^2 R^{-1}$, or $-R^5$.  Since $R^5$ grows faster than $R^3$, we can make the expected energy $\langle \psi, (E + V) \psi \rangle$ arbitrarily large and negative as $N,R \to \infty$.

Note the interesting contrast between this result and some previous ones we have seen.  In Newtonian mechanics, the energy of particles attracting each other with a $1/r$ potential is unbounded below.   In quantum mechanics, thanks the uncertainty principle, the energy is bounded below for any fixed number of particles.  However, quantum field theory allows for the creation of particles, and this changes everything!   Dyson's disaster arises because the vacuum can turn into a state with large numbers of electrons and positrons.  This disaster only occurs in an imaginary world where $\alpha$ is negative---but it may be enough to prevent the power series in QED from having a nonzero radius of convergence.  

We are left with a puzzle: how can perturbative QED work so well in practice, if the power series in QED diverge?

Much is known about this puzzle.   There is an extensive theory of `Borel summation', which allows one to extract well-defined answers from certain divergent power series. For example, consider a particle of mass $m$ on a line in a potential
\[    V(x) = x^2 + \beta x^4  .\]
When $\beta \ge 0$ this potential is bounded below, but when $\beta < 0$ it is not: classically, it describes a particle that can shoot to infinity in a finite time.   Let $H = K + V$ be the quantum Hamiltonian for this particle, where
\[  K = -\displaystyle{\frac{\hbar^2}{2m} \frac{\partial^2}{\partial x^2} } \]
is the usual operator for the kinetic energy and $V$ is the operator for potential energy.  When $\beta \ge 0$, the Hamiltonian $H$ is essentially self-adjoint on the set of smooth wavefunctions that vanish outside a bounded set.  Moreover, in this case $H$ has a `ground state': a state $\psi$ whose expected energy $\langle \psi, H \psi \rangle$ is as low as possible.   Call this expected energy $E(\beta)$.  One can show that $E(\beta)$ depends smoothly on $\beta$ for $\beta \ge 0$, and one can write down a Taylor series for $E(\beta)$.  

On the other hand, when $\beta < 0$, the Hamiltonian $H$ is \emph{not} essentially self-adjoint on the set of smooth wavefunctions that vanish outside a bounded interval.  This means that the quantum mechanics of a particle in this potential is ill-behaved when $\beta < 0$.  Heuristically speaking, the problem is that such a particle could tunnel through the barrier given by the local maxima of $V(x)$ and shoot off to infinity in a finite time.

This situation is similar to Dyson's disaster, since we have a theory that is well behaved for $\beta \ge 0$ and ill behaved for $\beta < 0$.  As before, the bad behavior seems to arise from our ability to convert an infinite amount of potential energy into other forms of energy.    However, in this simpler situation one can \emph{prove} that the Taylor series for $E(\beta)$ does not converge.  Barry Simon \cite{Simon1970} did this around 1969.  Moreover, one can prove that Borel summation, applied to this Taylor series, gives the correct value of $E(\beta)$ for $\beta \ge 0$ \cite{GGS}.  The same is known to be true for certain quantum field theories \cite{Rivasseau}.  Analyzing these examples, one can see why summing the first few terms of a power series can give a good approximation to the correct answer even though the series diverges.   The terms in the series get smaller and smaller for a while, but eventually they become huge.

Unfortunately, nobody has been able to carry out this kind of analysis for quantum electrodynamics.  In fact, the current conventional wisdom is that this theory is inconsistent, due to problems at very short distance scales.   In our discussion so far, we summed over Feynman diagrams with $\le n$ vertices to get the first $n$ terms of power series for answers to physical questions.  However, one can also sum over all diagrams with $\le n$ loops: that is, graphs with genus $\le n$.  This more sophisticated approach to renormalization, which sums over infinitely many diagrams, may dig a bit deeper into the problems faced by quantum field theories.   

If we use this alternate approach for QED we find something surprising.  Recall that
in renormalization we impose a momentum cutoff $\Lambda$, essentially ignoring waves of wavelength less than $\hbar/\Lambda$, and use this to work out a relation between the
the electron's bare charge $e_\bare(\Lambda)$ and its renormalized charge $e_\ren$.
We try to choose $e_\bare(\Lambda)$ that makes $e_\ren$ equal to the electron's experimentally observed charge $e$.   If we sum over Feynman diagrams with $\le n$ vertices this is always possible.  But if we sum over Feynman diagrams with at most one loop, it ceases to be possible when $\Lambda$ reaches a certain very large value, namely
\[     \Lambda \; = \; \exp\left(\frac{3 \pi}{2 \alpha} + \frac{5}{6}\right) m_e c \; \approx \; e^{647} m_e c. \] 
According to this one-loop calculation, the electron's bare charge becomes \emph{infinite} at this point!    This value of $\Lambda$ is known as a `Landau pole', since it was first noticed in about 1954 by Lev Landau and his colleagues \cite{Landau}.  

What is the meaning of the Landau pole?  I said that poetically speaking, the bare charge of the electron is the charge we would see if we could strip off the electron's virtual particle cloud.  A somewhat more precise statement is that $e_\bare(\Lambda)$ is the charge we would see if we collided two electrons head-on with a momentum on the order of $\Lambda$.  In this collision, there is a good chance that the electrons would come within a distance of $\hbar/\Lambda$ from each other.  The larger $\Lambda$ is, the smaller this distance is, and the more we penetrate past the effects of the virtual particle cloud, whose polarization `shields' the electron's charge. Thus, the larger $\Lambda$ is, the larger $e_\bare(\Lambda)$ becomes.  So far, all this makes good sense: physicists have done experiments to actually measure this effect.  The problem is that $e_\bare(\Lambda)$ becomes infinite when $\Lambda$ reaches a certain huge value. 

Of course, summing only over diagrams with $\le 1$ loops is not definitive.  Physicists have repeated the calculation summing over diagrams with $\le 2$ loops, and again found a Landau pole.  But again, this is not definitive.  Nobody knows what will happen as we consider diagrams with more and more loops.  Moreover, the distance $\hbar/\Lambda$ corresponding to the Landau pole is absurdly small!  For the one-loop calculation quoted above, 
this distance is about 
\[       e^{-647} \frac{\hbar}{m_e c} \; \approx \;  6 \times 10^{-294}\, \mathrm{meters} .\]
This is hundreds of orders of magnitude smaller than the length scales physicists have explored so far.  Currently the Large Hadron Collider can probe energies up to about 10 TeV, and thus distances down to about $2 \times 10^{-20}$ meters, or about 0.00002 times the radius of a proton.  Quantum field theory seems to be holding up very well so far, but no reasonable physicist would be willing to extrapolate this success down to $6 \times 10^{-294}$ meters, and few seem upset at problems that manifest themselves only at such a short distance scale.  

Indeed, attitudes on renormalization have changed significantly since 1948, when Feynman, Schwinger and Tomonoga developed it for QED.   At first it seemed a bit like a trick.  Later, as the success of renormalization became ever more thoroughly confirmed, it became accepted.  However, some of the most thoughtful physicists remained worried.   In 1975, Dirac said:
\begin{quote}
Most physicists are very satisfied with the situation. They say: `Quantum electrodynamics is a good theory and we do not have to worry about it any more.' I must say that I am very dissatisfied with the situation, because this so-called `good theory' does involve neglecting infinities which appear in its equations, neglecting them in an arbitrary way. This is just not sensible mathematics. Sensible mathematics involves neglecting a quantity when it is small---not neglecting it just because it is infinitely great and you do not want it!
\end{quote}
As late as 1985, Feynman wrote:
\begin{quote}
The shell game that we play [...] is technically called `renormalization'. But no matter how clever the word, it is still what I would call a dippy process! Having to resort to such hocus-pocus has prevented us from proving that the theory of quantum electrodynamics is mathematically self-consistent. It's surprising that the theory still hasn't been proved self-consistent one way or the other by now; I suspect that renormalization is not mathematically legitimate.
\end{quote}
By now renormalization is thoroughly accepted among physicists.  The key move was a change of attitude emphasized by Kenneth Wilson in the 1970s \cite{WilsonKogut}.  Instead of treating quantum field theory as the correct description of physics at arbitrarily large energy-momenta, we can assume it is only an approximation.  For renormalizable theories, one can argue that even if quantum field theory is inaccurate at large energy-momenta, the corrections become negligible at smaller, experimentally accessible energy-momenta.   If so, instead of seeking to take the $\Lambda \to \infty$ limit, we can use renormalization to relate bare quantities at some large but finite value of $\Lambda$ to experimentally observed quantities.

From this practical-minded viewpoint, the possibility of a Landau pole in QED is less important than the behavior of the Standard Model.  Physicists believe that the Standard Model would suffer from Landau pole at momenta low enough to cause serious problems if the Higgs boson were considerably more massive than it actually is.  Thus, they were relieved when the Higgs was discovered at the Large Hadron Collider with a mass of about $125 \textrm{ GeV}/c^2$.  However, the Standard Model may still suffer from a Landau pole at high momenta, as well as an instability of the vacuum \cite{JKK}.  

Regardless of practicalities, for the \emph{mathematical} physicist, the question of whether or not QED and the Standard Model can be made into well-defined mathematical structures that obey the axioms of quantum field theory remain open problems of great significance.   Most physicists believe that this can be done for pure Yang--Mills theory, but actually proving this is the first step towards winning \$1,000,000 from the Clay Mathematics Institute.

\section{General relativity}

Combining electromagnetism with relativity and quantum mechanics led to QED, and we have seen the immense struggles with the continuum this caused.   Combining gravity with relativity led Einstein to \emph{general relativity}.   

In general relativity, infinities coming from the continuum nature of spacetime are deeply connected to its most dramatic successful predictions: black holes and the Big Bang.   In this theory, the density of the Universe approaches infinity as we go back in time toward the Big Bang, and the density of a star approaches infinity as it collapses to form a black hole.   Thus we might say that instead of struggling against infinities, general relativity \emph{accepts} them and has learned to live with them. 

General relativity does not take quantum mechanics into account, so the story is not  yet over.   Many physicists hope that quantum gravity will eventually save physics from its struggles with the continuum.   Simple dimensional analysis suggests that quantum gravity effects may become important at length scales near the `Planck length':
\[                  \ell_p = \sqrt{\frac{\hbar G}{c^3}} \approx 1.6 \times 10^{-35} \textrm{ meters}  .\]
Unfortunately, this is too small for direct experiments at present.  The hope that something new happens around this length scale has motivated a profusion of new ideas on spacetime: too many to survey here.  Instead, I shall focus on the humbler issue of how singularities arise in general relativity---and why they might not rob this theory of its predictive power.

General relativity says that spacetime is a 4-dimensional Lorentzian manifold. Thus, it can be covered by patches equipped with coordinates, so that in each patch we can describe points by lists of four numbers.  Any curve $\gamma(s)$ going through a point then has a tangent vector $v$ whose components are $v^\mu = d \gamma^\mu(s)/ds$.    Furthermore, given two tangent vectors $v,w$ at the same point we 
can take their inner product
\[       g(v,w) = g_{\mu \nu} v^\mu w^\nu  \]
where as usual we sum over repeated indices, and $g_{\mu \nu}$ is a $4 \times 4$ matrix called the metric, depending smoothly on the point.    We require that at any point we can find some coordinate system where this matrix takes the usual Minkowski form:
\[   g =  \left( \begin{array}{cccc} 
                 -1 & 0 &0 & 0  \\ 
                 0     & 1 &0 & 0  \\
                 0     & 0 &1 & 0  \\
                 0     & 0 &0 & 1  \\
\end{array}\right).
\]
However, as soon as we move away from our chosen point, the form of the matrix $g$ in these particular coordinates may change.   

General relativity says how the metric is affected by matter.   It does this in a single equation, Einstein's equation, which relates the `curvature' of the metric at any point to the flow of energy and momentum through that point.   

To work with the concept of curvature, Einstein had to learn differential geometry from his mathematician friend Marcel Grossman.    One of the great delights of general relativity is how much more can be rigorously proved about this theory than quantum field theory, where even the basic formalism remains problematic.  The price to pay is a lot of differential geometry.   Instead of explaining all this, I will take some shortcuts and focus on providing intuition.    It helps to reformulate Einstein's equation in terms of the motion of particles.  For more details, and a list of resources for further study, see \cite{BaezBunn}.

To understand Einstein's equation, let us see what it says about a small round ball of test particles that are initially all at rest relative to each other.   The scenario here requires a bit of explanation.   First, because spacetime is curved, it only looks like Minkowski spacetime---the world of special relativity---in the limit of a very small region.   The concepts of `round' and `at rest relative to each other' only make sense in this limit.  Thus, the forthcoming statement of Einstein's equation is precise only in this limit.  Of course, taking this limit relies on the fact that spacetime is a continuum.

Second, a `test particle' is a classical point particle with so little mass that while it is affected by gravity, its effects on the geometry of spacetime are negligible.  We assume our test particles are affected only by gravity, no other forces.   In general relativity this means that they move along timelike geodesics.   Roughly speaking, these are paths that go slower than light and bend as little as possible.   We can make this precise without much work.

For a path in \emph{space} to be a geodesic means that if we slightly vary any small portion of it, it can only become longer.  However, a path $\gamma(s)$ in \emph{spacetime} traced out by particle moving slower than light must be `timelike', meaning that its tangent vector $v = \gamma'(s)$ satisfies $g(v,v) < 0$.   We define the proper time along such a path from $s = s_0$ to $s = s_1$ to be 
\[               \int_{s_0}^{s_1} \sqrt{-g(\gamma'(s),\gamma'(s))} \, ds .\]
This is the time ticked out by a clock moving along that path.   A timelike path is a geodesic if the proper time can only \emph{decrease} when we slightly vary any small portion of it.   Particle physicists prefer the opposite sign convention for the metric, and then we do not need the minus sign under the square root.  But the fact remains the same: timelike geodesics locally maximize the proper time.

Actual particles are not test particles.  First, the concept of test particle does not take quantum theory into account.  Second, all known particles are affected by forces other than gravity.  Third, any actual particle affects the geometry of the spacetime it inhabits.   Test particles are just a mathematical trick for studying the geometry of spacetime.  Still, a sufficiently light particle that is affected very little by forces other than gravity should be well approximated by a test particle, though rigorously proving this is difficult \cite{EIH}.  For example, an artificial satellite moving through the Solar System behaves like a test particle if we ignore the solar wind, the radiation pressure of the Sun, and so on.  
 
If we start with a small round ball consisting of many test particles that are initially all at rest relative to each other, to first order in time it will not change shape or size.  However, to second order in time it can expand or shrink, due to the curvature of spacetime.
It may also be stretched or squashed, becoming an ellipsoid.  This should not be too surprising, because any linear transformation applied to a ball gives an ellipsoid.  

Let $V(t)$ be the volume of the ball after a time $t$ has elapsed, where time is measured by a clock attached to the particle at the center of the ball.  Then in units where $c = G = 1$, Einstein's equation says: 
\[ \left.{\ddot V\over V} \right|_{t = 0} =
 -4\pi \left( \begin{array}{l} 
{\rm flow \; of \;} t{\rm -momentum \; in \; the \;\,} t {\rm \,\; direction \;} + \\ 
{\rm flow \; of \;} x{\rm -momentum \; in \; the \;\,} x {\rm \; direction \;} + \\ 
{\rm flow \; of \;} y{\rm -momentum \; in \; the \;\,} y {\rm \; direction \;} + \\ 
{\rm flow \; of \;} z{\rm -momentum \; in \; the \;\,} z {\rm \; direction} 
\end{array} \right) .\] 
These flows here are measured at the center of the ball at time zero, and the coordinates used here take advantage of the fact that to first order, at any one point, spacetime looks like Minkowski spacetime.  

The flows in Einstein's equation are the diagonal components of a $4 \times 4$ matrix $T$ called the `stress-energy tensor'.  The components $T_{\alpha \beta}$ of this matrix say how much momentum in the $\alpha$ direction is flowing in the $\beta$ direction through a given point of spacetime.   Here $\alpha$ and $\beta$ range from $0$ to $3$, corresponding to the $t,x,y$ and $z$ coordinates.  For example, $T_{00}$ is the flow of $t$-momentum in the $t$-direction.  This is just the energy density, usually denoted $\rho$.  The flow of $x$-momentum in the $x$-direction is the pressure in the $x$ direction, denoted $P_x$, and similarly for $y$ and $z$.   The reader may be more familiar with direction-independent pressures, but it is easy to manufacture a situation where the pressure depends on the direction: just squeeze a book between one's hands.   

Thus, Einstein's equation says
\[   {\ddot V\over V} \Bigr|_{t = 0} 
= -4\pi (\rho + P_x + P_y + P_z).  \]
It follows that positive energy density and positive pressure both curve spacetime in a way that makes a freely falling ball of point particles tend to shrink.  Since $E = mc^2$ and we are working in units where $c = 1$, ordinary mass density counts as a form of energy density.  Thus a massive object will make a swarm of freely falling particles at rest around it start to shrink.  In short, \emph{gravity attracts}.

Already from this, gravity seems dangerously inclined to create singularities.  Suppose that instead of test particles we start with a stationary cloud of `dust': a fluid of particles having nonzero energy density but no pressure, moving under the influence of gravity alone.  The dust particles will still follow geodesics, but they will affect the geometry of spacetime.  Their energy density will make the ball start to shrink.  As it does, the energy density $\rho$ will increase, so the ball will tend to shrink ever faster, approaching infinite density in a finite amount of time.   This in turn makes the curvature of spacetime become infinite in a finite amount of time.  The result is a `singularity'. 

In reality, matter is affected by forces other than gravity.  Repulsive forces may prevent gravitational collapse.  However, this repulsion creates pressure, and Einstein's equation says that pressure also creates gravitational attraction!   In some circumstances this can overwhelm whatever repulsive forces are present.  Then the matter collapses, leading to a singularity---at least according to general relativity.

When a star more than 8 times the mass of our Sun runs out of fuel, its core suddenly collapses.  The surface is thrown off explosively in an event called a supernova.   Most of the energy---the equivalent of thousands of Earth masses---is released in a ten-second burst of neutrinos, formed as a byproduct when protons and electrons combine to form neutrons.  If the star's mass is below 20 times that of our the Sun, its core crushes down to a large ball of neutrons with a crust of iron and other elements: a neutron star.

However, this ball is unstable if its mass exceeds the Tolman--Oppenheimer--Volkoff limit, somewhere between 1.5 and 3 times that of our Sun.   Above this limit, gravity overwhelms the repulsive forces that hold up the neutron star.  And indeed, no neutron stars heavier than 3 solar masses have been observed.   Thus, for very heavy stars, the endpoint of collapse is not a neutron star, but something else: a \emph{black hole}, an object that bends spacetime so much even light cannot escape.

If general relativity is correct, a black hole contains a singularity.  Many physicists expect that general relativity breaks down inside a black hole, perhaps because of quantum effects that become important at strong gravitational fields.  The singularity is considered a strong hint that this breakdown occurs.   If so, the singularity may be a purely theoretical entity, not a real-world phenomenon.  Nonetheless, everything we have observed about black holes matches what general relativity predicts.  Thus, unlike all the other theories we have discussed, general relativity predicts infinities that are connected to striking phenomena that are \emph{actually observed}.

The Tolman--Oppenheimer--Volkoff limit is not precisely known, because it depends on properties of nuclear matter that are not well understood \cite{Bombaci}.  However, there are theorems that say singularities \emph{must} occur in general relativity under certain conditions.  

One of the first was proved by Raychauduri \cite{Raychauduri} and Komar \cite{Komar}
in the mid-1950's.   It applies only to  `dust', and indeed it is a precise version of our verbal argument above.   It introduced the `Raychauduri equation', which is the geometrical way of thinking about spacetime curvature as affecting the motion of a small ball of test particles.   It shows that under suitable conditions, the energy density must approach infinity in a finite amount of time along the path traced out out by a dust particle.   

The first required condition is that the flow of dust be initally converging, not expanding.  The second condition, not mentioned in our verbal argument, is that the dust be `irrotational', not swirling around.     The third condition is that the dust particles be affected only by gravity, so that they move along geodesics.   Due to the last two conditions, the Raychauduri--Komar theorem does not apply to collapsing stars.

The more modern singularity theorems eliminate these conditions.  But they do so at a price: they require a more subtle concept of singularity.   There are various possible ways to define this concept.  They are all a bit tricky, because a singularity is not a point or region in spacetime.  

For our present purposes, we shall define a singularity to be an `incomplete timelike or null geodesic'.    As already explained, a timelike geodesic is the kind of path traced out by a test  particle moving slower than light.  Similarly, a null geodesic is the kind of path traced out by a test particle moving at the speed of light.   We say a geodesic is `incomplete' if it ceases to be well-defined after a finite amount of time.   For example, general relativity says a test particle falling into a black hole follows an incomplete geodesic.   In a rough-and-ready way, people say the particle `hits the singularity'.   But the singularity is not a place in spacetime.  What we really mean is that the particle's path becomes undefined after a finite amount of time.

We need to be a bit careful about the role of `time' here.  For test particles moving slower than light this is easy, since we can parametrize a timelike geodesic by proper time.  However, the tangent vector $v = \gamma'(s)$ of a null geodesic has $g(v,v) = 0$, so a particle moving along a null geodesic does not experience any passage of proper time.  Still, any geodesic, even a null one, has a family of preferred parametrizations.  These differ only by reparametrizations of the form $s \mapsto as + b$.   By `time' we really mean the variable $s$ in any of these preferred parametrizations.   Thus, if our spacetime is some Lorentzian manifold $M$, we say a geodesic $\gamma \maps [s_0, s_1] \to M$ is incomplete if, parametrized in one of these preferred ways, it cannot be extended to a strictly longer interval.  

The first modern singularity theorem was proved by Penrose \cite{Penrose1965} in 1965.  It says that if space is infinite in extent, and light becomes trapped inside some bounded region, and no exotic matter is present to save the day, either a singularity or something even more bizarre must occur.   This theorem applies to collapsing stars.   When a star of sufficient mass collapses, general relativity says that its gravity becomes so strong that light becomes trapped inside some bounded region.   We can then use Penrose's theorem to analyze the possibilities.

Shortly thereafter Hawking proved a second singularity theorem, which applies to the Big Bang \cite{Hawking1966}.   It says that if space is finite in extent, and no exotic matter is present, generically either a singularity or something even more bizarre must occur.  The singularity here could be either a Big Bang in the past, a Big Crunch in the future, both---or possibly something else.  Hawking also proved a version of his theorem that applies to certain Lorentzian manifolds where space is infinite in extent, as seems to be the case in our Universe.  This version requires extra conditions.

There are some undefined phrases in this summary of the Penrose--Hawking singularity theorems, most notably these:
\begin{itemize}
\item `exotic matter'
\item `singularity'
\item `something even more bizarre'.
\end{itemize}
So, let me say a bit about each.

These theorems precisely specify what is meant by `exotic matter'.  All known forms of matter obey the `dominant energy condition', which says that
\[       |P_x|, \, |P_y|, \, |P_z| \le \rho \]
at all points and in all locally Minkowskian coordinates.  Exotic matter is anything that violates this condition.  For a detailed discussion of this and other energy conditions, see the survey by Curiel \cite{Curiel}.

The Penrose--Hawking singularity theorems also say what counts as `something even more bizarre'.  An example would be a closed timelike curve.   A particle following such a path would move slower than light yet eventually reach the same point where it started: and not just the same point in space, but the same point in \emph{spacetime}.  If you could do this perhaps you could wait, see if it rains tomorrow, and then go back in time and decide whether to buy an umbrella today.  There are certainly solutions of Einstein's equation with closed timelike curves.  The first interesting one was found by Einstein's friend G\"odel in 1949, as part of an attempt to probe the nature of time \cite{Goedel}.  However, closed timelike curves are generally considered less plausible than singularities.

In the Penrose--Hawking singularity theorems, `something even more bizarre' means that spacetime is not `globally hyperbolic'.  To understand this, we need to think about when we can predict the future or past given initial data.  When studying field equations like Maxwell's theory of electromagnetism or Einstein's theory of gravity, physicists like to specify initial data on space at a given moment of time.  However, in general relativity there is considerable freedom in how we choose a slice of spacetime and call it `space'.   What should we require?   For starters, we want a 3-dimensional submanifold $S$ of spacetime that is `spacelike': every vector $v$ tangent to $S$ should have $g(v,v) > 0$.   However, we also want any timelike or null curve to hit $S$ exactly once.  A spacelike surface with this property is called a `Cauchy surface', and a Lorentzian manifold containing a Cauchy surface is said to be `globally hyperbolic'.   Globally hyperbolicity excludes closed timelike curves, but also other bizarre behavior.  In a globally hyperbolic spacetime, we can predict the future or past given initial data on any Cauchy surface.  

By now the original singularity theorems have been greatly generalized and clarified.  Hawking and Penrose \cite{HawkingPenrose1970} gave a unified treatment of both theorems in 1970.  The textbook by Hawking and Ellis \cite{HawkingEllis} gave the first
really systematic introduction to this subject, and Wald's text gives a shorter, 
more modern treatment \cite{Wald1984}.  Hawking's 1994 lectures \cite{Hawking1994} give a beautiful overview of the key ideas, and a paper by Garfinkle and Senovilla \cite{Hawking1994} reviews the subject and its history up to 2015.   

If we accept that general relativity really predicts the existence of singularities in physically realistic situations, the next step is to ask whether they rob general relativity of its predictive power.  The `cosmic censorship hypothesis', proposed by Penrose in 1969, claims they do not \cite{Penrose1969}.   

To formulate such a conjecture, we must first think about what behaviors we consider acceptable.  Consider first a black hole formed by the collapse of a star.  According to general relativity, matter can fall into this black hole and `hit the singularity' in a finite amount of proper time, but nothing can come out of the singularity.  The time-reversed version of a black hole, called a `white hole', is often considered more disturbing.  White holes have never been seen, but they are mathematically valid solutions of Einstein's equation.  In a white hole, matter can come \emph{out} of the singularity, but nothing can fall \emph{in}.    Naively, this seems to imply that the future is unpredictable given knowledge of the past.   Of course, the same logic applied to black holes would say the past is unpredictable given knowledge of the future.

If white holes are disturbing, perhaps the Big Bang should be more so.  In the usual solutions of general relativity describing Big Bang cosmologies, \emph{all matter in the universe} comes out of a singularity!  More precisely, if one follows any timelike geodesic back into the past, it becomes undefined after a finite amount of proper time.  Naively, this may seem a massive violation of predictability: in this scenario, the whole universe `sprang out of nothing' about 14 billion years ago. 

However, in all three examples so far---astrophysical black holes, their time-reversed versions and the Big Bang---spacetime is globally hyperbolic.  Thus, we can specify data on a Cauchy surface and solve Einstein's equations to predict the future (and past) development of the metric throughout all of spacetime.   How is this compatible with the naive intuition that a singularity causes a failure of predictability?   

For any globally hyperbolic spacetime $M$, one can find a smoothly varying family of Cauchy surfaces $S_t$ ($t \in \R$) such that each point of $M$ lies on exactly one of these surfaces.  This amounts to a way of chopping spacetime into `slices of space' for various choices of the `time' parameter $t$.  For an astrophysical black hole, the singularity is in the future of all these surfaces.   That is, a timelike or null geodesic that hits the singularity must go through all the surfaces $S_t$ before it becomes undefined.  Similarly, for a white hole or the Big Bang, the singularity is in the past of all these surfaces.  In either case, the singularity cannot interfere with our predictions of what occurs in spacetime.  For more on this topic, try Earman's delightful book \textsl{Bangs, Crunches, Whimpers and Shrieks: Singularities and Acausalities in Relativistic Spacetimes} \cite{Earman}.

A more challenging example is posed by the Kerr--Newman solution of Einstein's equation
coupled to the vacuum Maxwell equations \cite{Wald1984}.     When 
\[         e^2 + (J/m)^2 < m^2 ,\]
this solution describes a rotating charged black hole with mass $m$, charge $e$ and angular momentum $J$ in units where $c = G = 1$.    In 1968, 
Carter \cite{Carter1968} noted that the Kerr--Newman solution acts like a particle with gyromagnetic ratio $g = 2$, surprisingly close to that of an electron.  However, an electron has
\[         e^2 + (J/m)^2 \gg m^2 .\]
The Kerr--Newman solution still has $g = 2$ in this case, but also some
disturbing pathological features.   It has closed timelike curves accessible from
the outside world!  It also has a `naked singularity'.  Roughly speaking, this is a singularity that can be seen by arbitrarily faraway observers in a spacetime whose geometry asymptotically approaches that of Minkowski spacetime.  A spacetime with a naked singularity cannot be globally hyperbolic \cite{HawkingEllis}.    

The cosmic censorship hypothesis comes in a number of forms.    The original version due to Penrose is now called `weak cosmic censorship' \cite{Wald2008}.  It asserts that
in a spacetime whose geometry asymptotically approaches that of Minkowski spacetime, gravitational collapse cannot produce a naked singularity.   

In 1991, Preskill and Thorne made a bet against Hawking in which they claimed that weak cosmic censorship was false.  Hawking conceded this bet in 1997 when a counterexample was found.  This features finely-tuned infalling matter poised right on the brink of forming a black hole.   It \emph{almost} creates a region from which light cannot escape---but not quite.  Instead, it creates a naked singularity!

Given the delicate nature of this construction, Hawking did not give up.  Instead
he made a second bet, which says that weak cosmic censorshop holds `generically'---
that is, for an open dense set of initial conditions.  In 1999, Christodoulou proved that for spherically symmetric solutions of Einstein's equation coupled to a massless scalar field, weak cosmic censorship holds generically \cite{Christodoulou1999}.  While spherical symmetry is a very restrictive assumption, this result is a good example of how, with plenty of work, we can make progress in rigorously settling the questions raised by general relativity.

Indeed, Christodoulou has been a leader in this area.   For example, the vacuum Einstein equations have solutions describing gravitational waves, much as the vacuum Maxwell equations have solutions describing electromagnetic waves.  However, gravitational waves can actually form black holes when they collide.  This raises the question of the stability of Minkowski spacetime.  Must sufficiently small perturbations of the Minkowski metric go away in the form of gravitational radiation, or can tiny wrinkles in the fabric of spacetime somehow amplify themselves and cause trouble---perhaps even a singularity?   In 1993, together with Klainerman, Christodoulou proved that Minkowski spacetime is indeed stable \cite{CK}.  Their proof fills a 514-page book.

In 2008, Christodoulou completed an even longer rigorous study of the formation of black holes \cite{Christodoulou2008}.  This can be seen as a vastly more detailed look at questions which Penrose's original singularity theorem addressed in a general, preliminary way.  Nonetheless, there is much left to be done to understand the behavior of singularities in general relativity \cite{Rendall}.

\section{Conclusions}

We have seen that in every major theory of physics, challenging mathematical 
questions arise from the assumption that spacetime is a continuum.  The continuum threatens us with infinities.  Do these infinities threaten our ability to extract predictions from these theories---or even our ability to formulate these theories in a precise way?   We can answer these questions, but only with hard work.   Is this a sign that we are somehow on the wrong track?  Is the continuum as we understand it only an approximation to some deeper model of spacetime?   Only time will tell.   Nature is providing us with plenty of clues, but it will take patience to read them correctly.

\subsection*{Acknowledgements}

I thank Flip Tancredo for his Feynman diagram package, and Emory Bunn and Greg Weeks for helpful comments.

\end{document}